\documentclass[prl,twocolumn,superscriptaddress,preprintnumbers,floatfix]{revtex4-2}
\usepackage[T1]{fontenc}
\usepackage{epsfig}
\usepackage{graphicx}
\usepackage[english]{babel}
\usepackage{hyphenat}
\usepackage{amsmath}
\usepackage{amssymb}
\usepackage{mathtools}
\usepackage{mathrsfs}
\usepackage{slashed}
\usepackage{epstopdf}
\usepackage{xcolor}
\usepackage{braket}
\usepackage{booktabs}
\definecolor{lcolor}{rgb}{0.,0.0,0.}
\definecolor{citcolor}{rgb}{0,0.,0.5}
\usepackage[breaklinks,colorlinks,urlcolor=blue,citecolor=blue,linkcolor=blue]{hyperref}
\usepackage{multirow}
\usepackage{ltablex}
\usepackage{soul}

\usepackage[e]{esvect}

\newcommand{\secn}[1]{Section~1}
\newcommand{\appn}[1]{Appendix~1}

\long\def\comment#1{ }

\def\and{\quad\text{and}\quad}

\def\0{{\boldsymbol 0}}
\def\1{{\boldsymbol 1}}
\def\p{{\boldsymbol p}}

\def\0{{\boldsymbol 0}}

\renewcommand\d{\delta}

\newcommand\s{\sigma}

\newcommand{\tvec}{\boldsymbol}

\renewcommand{\part}{{\rm part}}

\newcommand{\be}{\begin{equation}}
\newcommand{\ee}{\end{equation}}
\newcommand{\bes}{\begin{subequations}}
\newcommand{\ees}{\end{subequations}}
\newcommand{\bea}{\begin{eqnarray}}
\newcommand{\eea}{\end{eqnarray}}

\newcommand{\pa}{\partial}

\def\bea#1\eea{\begin{align}#1\end{align}}
\newcommand{\bef}{\begin{figure}[h!tb]\centering}
\newcommand{\eef}{\end{figure}}

\allowdisplaybreaks

\newcommand{\n}{{\bf n}}

\begin{document}

\preprint{CERN-TH-2026-162} 

\title{Anomaly Realization in Charge-Flux Detector Correlators}

\author{Jo\~{a}o Barata}
\email{joao.lourenco.henriques.barata@cern.ch}
\affiliation{CERN, Theoretical Physics Department, CH-1211, Geneva 23, Switzerland}

\author{Ratmir Jumanov}
\email{dzhumanov.r19@physics.msu.ru}
\affiliation{ Moscow State University, Faculty of Physics, Department  of Theoretical Physics, 119991, Moscow, Russia}

\author{Andrey V. Sadofyev}
\email{andrey.sadofyev@ehu.eus}
\affiliation{Department of Physics, University of the Basque Country UPV/EHU, P.O. Box 644, 48080 Bilbao, Spain}
\affiliation{IKERBASQUE, Basque Foundation for Science, Plaza Euskadi 5, 48009 Bilbao, Spain}

\begin{abstract}
Quantum anomalies provide a bridge between ultraviolet properties of a theory and its infrared sector. We study how this connection appears in axial-charge-flow observables. In the simplest example, an axial-charge detector probes the fermionic cut of the anomalous triangle and resolves its infrared content as an angular distribution. The massless limit does not commute with the angular integration: a contribution suppressed at fixed angle collapses onto the two beam-collinear directions while retaining the finite integrated sum rule fixed by the axial anomaly. We then replace the axial-charge detector by higher-spin helicity (zilch) detectors and study a family of axial-anomaly-controlled energy-weighted sum rules for the corresponding fluxes. We further show that the same singular localization mechanism and finite zilch-flux sum rules persist in the mixed axial-gravitational channel. We briefly comment on extensions to more general states and multipoint correlators.
\end{abstract}

\maketitle

\noindent\textbf{Introduction:} Quantum anomalies are exact quantum phenomena, rooted in the path-integral measure, yet accessible with the standard toolkit of perturbative quantum field theory \cite{Bertlbook}. They consist of the violation of a classical symmetry by quantum effects, connecting the ultraviolet (UV) and infrared (IR) sectors of the underlying theory, see \cite{Giannotti:2008cv} for a detailed discussion. Their IR realization involves massless “Dolgov-Zakharov” poles \cite{Dolgov:1971ri,Frishman:1980dq,Coleman:1982yg} and may lead to macroscopic manifestations in systems ranging from the quark-gluon plasma and topological materials, to astrophysical objects and the primordial plasma of the early universe, see e.g. \cite{Kharzeev:2015znc,Huang:2015oca,Landsteiner:2016led,Kamada:2022nyt} for a review. Apart from their practical importance, anomalies provide one of the most elegant examples of the interplay between quantum consistency and observable long-distance physics.

In the context of collider physics, it is natural to ask how quantum anomalies manifest themselves at the level of detector based observables. 
Recent developments have established energy-flow detector operators as a natural quantum field theory language for characterizing collider final states \cite{Hofman:2008ar,Belitsky:2013bja,Belitsky:2014zha,Hartman:2016lgu,Hartman:2023qdn}, for a recent review see \cite{Moult:2025nhu}. 
These admit an operator formulation in terms of asymptotic fluxes \cite{Basham:1978zq,Basham:1979gh,Basham:1977iq,Basham:1978bw,Ore:1979ry,Sveshnikov:1995vi,Tkachov:1995kk,Korchemsky:1999kt}, and can be directly connected to their experimental realization. More, this framework admits straightforward generalizations to detectors of other conserved charges, such as electric charge or baryon number, see e.g. \cite{Hofman:2008ar,Belitsky:2013bja,Belitsky:2014zha,Chicherin:2020azt,Monni:2025zyv}. Thus the primary goal of this work is to clarify how quantum anomalies are realized through these detector operators.

Throughout this work we will mainly focus on the axial anomaly \cite{Adler:1969gk,Bell:1969ts}, consisting of the violation of the axial current of massless fermions in background electromagnetic (EM) fields. The same detector-based logic can be extended, upon choosing the appropriate detector, to other classes of anomalies, such as the scale anomaly \cite{Crewther:1972kn,Adler:1976zt,Collins:1976yq,Nielsen:1977sy,Duff:1977ay,Bhattacharya:2022xxw}. The elementary detector operator we consider is
\begin{equation}
\mathcal{Q}_{5}(\n)=\lim_{R\rightarrow\infty}R^{2}\int_{0}^{\infty}dt\,n_{i}J_{5}^{i}(t,R\n)\, .
\label{Q5def}
\end{equation}
We shall refer to it as the axial-flux (detector) operator. The corresponding elementary observable, $\Sigma_5(\n)=\langle \Psi|\mathcal{Q}_{5}(\n)|\Psi\rangle$, defined on a particular initial (normalized) state $|\Psi\rangle $, obeys an anomalous operator sum rule upon angular integration.

\vspace{-0.2cm}

\begin{figure}[h!]
    \centering
    \includegraphics[width=.5\columnwidth]{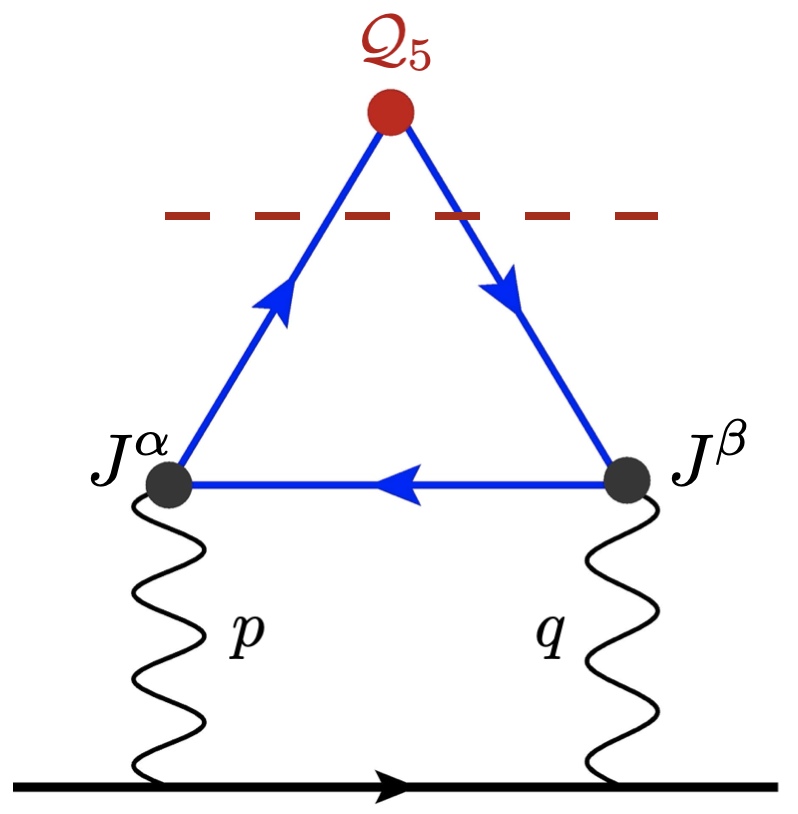}
    \vspace{-0.25cm}
    \caption{Anomalous triangle embedded in an $e^+e^-$--initiated process. The dashed line schematically indicates the cut through the on-shell fermionic state resolved by $\mathcal{Q}_5$.}
    \label{triangle}
\end{figure}

\vspace{-0.25cm}

The axial anomaly is often introduced through its UV face: the regularization of the corresponding triangle diagram. However, it also possesses an IR face, encoded in the spectral representation of the same triangle. In particular, the anomalous sum rule can be saturated by a massless pole emerging from a singular limit of massive intermediate states, see e.g. \cite{Giannotti:2008cv,Armillis:2009im,Kirilin:2013fqa,Mottola:2019nui,Tarasov:2020cwl,Tarasov:2021yll,Mottola:2023emy,Coriano:2023hts,Coriano:2024ive}. The axial-flux detector makes this IR realization directly visible by resolving one of the possible cuts of the triangle graph in angle. Here the insertion of the $\mathcal{Q}_5$ detector and the two on-shell fermion trajectories form the same triangle kinematics as the anomalous amplitude, although the cut resolved by the detector differs from its linear discontinuity in the spectral realization of the anomaly.

Probably the best-known experimental manifestation of the anomaly is the decay $\pi^0\to 2\gamma$, although its interpretation requires matching the quark-level anomaly onto the hadronic pion state \cite{Bertlbook, DonGolHol_SM}. Here we set aside this non-perturbative matching and isolate the anomalous mechanism in a fully perturbative setting. In Fig.~\ref{triangle} we illustrate its embedding into an $e^+e^-$--initiated process: the incoming electron-positron state provides the EM source coupled to the triangle, while $\mathcal{Q}_5$ resolves the axial-flux carried by the on-shell fermions. While $\mathcal{Q}_5$ provides the cleanest theoretical realization, a crossed electroweak assignment can move the axial insertion to the source and probe the same anomalous amplitude through the electric-charge detector $\mathcal{Q}$. The recent revival of energy-correlator measurements in $e^+e^-$ using archival ALEPH data \cite{Electron-PositronAlliance:2025fhk} makes anomaly-sensitive extensions of this program particularly timely.

In what follows, we show that the cut of the anomalous triangle, corresponding to the axial-flux one-point function, can have a singular massless limit in the states that excite the relevant anomalous channel. For such states, at every fixed non-collinear angle, $|n_z|<1$, the mass-suppressed contribution responsible for the anomalous sum rule vanishes as the fermion mass is taken to zero, while its angular integral remains finite:
\begin{equation}
\lim_{m\rightarrow 0}\int_{\n}\Sigma_5(\n)\neq0\,.
\end{equation}
This behavior closely follows the logic of the spectral representation of the axial anomaly \cite{Dolgov:1971ri}, although the resolved cut is different. The resulting contribution is localized at the two beam-collinear endpoints and provides an angular detector counterpart of the spectral collapse that produces the anomaly pole. We further show that upon changing from the axial-flux detector to higher-spin helicity detectors, often referred to as zilches \cite{lipkin,kibble}, one finds a novel family of energy-weighted anomaly relations. We then extend the same construction to the mixed axial-gravitational channel and show that the singular localization mechanism and finite normalized zilch-flux sum rules persist. This structure is naturally suggested by earlier results on zilch vortical effects \cite{Chernodub:2018era, Huang:2020kik,Alexandrov:2020zsj,Hattori:2020gqh}, vortical transport of higher-spin chiral particles \cite{Avkhadiev:2017fxj,Yamamoto:2017uul,Hayata:2017tbr,Huang:2018aly,Hattori:2020gqh,Prokhorov:2020npf}, and the general structure of chiral effects \cite{Loganayagam:2012pz,Banerjee:2012cr,Dwivedi:2016kkj,Stone:2018zel}.

\noindent\textbf{Anomalous sum rule:}  At the operator level the axial anomaly divergence for a Dirac fermion takes the form
\begin{equation}
  \partial_{\mu}J_{5}^{\mu}
  =
  2im\,\bar\psi\gamma_{5}\psi
  +
  \frac{2\alpha}{\pi}\tvec{E}\cdot\tvec{B}\,.
  \label{eq:axial_anomaly}
\end{equation}
Using this Ward identity, we can readily constrain the outgoing axial-flux
\begin{equation}
\label{Q5sumrule}
  \int_{\n}\,
  \mathcal{Q}_{5}(\n)
  = \int_0^\infty dt\int dV\,\partial_{\mu}J_{5}^{\mu}-\Delta N_5\,,
\end{equation}
where $\Delta N_5=N_5(t)|^{\infty}_{0}$ is the operator corresponding to the change in the axial charge over the system evolution and the spatial integration goes over the detector volume. Because Eq.~\eqref{Q5sumrule} is an operator identity, it holds in any state and inside products with additional detector operators. It constrains the angle-integrated flux independently of the nature of its carriers, while its angular and energetic distribution remains state dependent.

Focusing on $\Sigma_5(\n)$, it is instructive to resolve the final state into asymptotic particle states. Inserting a complete set of outgoing
states $|X\rangle$ gives schematically
\begin{equation}
\label{Q5generalX}
  \Sigma_{5}(\n)
  =
  \frac{1}{\sigma_{\rm tot}}\sum_{X}
  \left|
    \mathcal{M}_{\Psi\rightarrow  X}
  \right|^{2}
  q_{5,X}(\n),
\end{equation}
where $q_{5,X}(\n)$ is the axial-flow weight defined by $\mathcal{Q}_5(\n)|X\rangle=q_{5,X}(\n)|X\rangle$, $\mathcal{M}_{\Psi\rightarrow  X}=\langle X|U(\infty,0)|\Psi\rangle$, and $\sigma_{\rm tot}=\sum_{X}\left|\mathcal{M}_{\Psi\rightarrow  X }\right|^{2}$. Here and below, $\sum_X$ includes the integration over the on-shell phase space and the sum over the relevant discrete quantum numbers. The same on-shell channel appears in the anomalous triangle discontinuity, to which we return below.

For a massive fermion of helicity $\sigma=\pm1$,
$\mathcal{Q}_{5}(\n)|\p,\sigma\rangle=\frac{\sigma}{\beta_{p}}\,\delta^{(2)}(\n-\n_p)|\p,\sigma\rangle$ with $\n_p=\p/|\p|$ and $\beta_{p}=|\mathbf p|/E_{p}$. The simplest initial state saturating the right-hand side of Eq.~\eqref{Q5sumrule} is a polarized two-photon source with center-of-mass energy $\sqrt{s}$. At the leading order it scatters into a fermion-antifermion pair:
\begin{equation}
\gamma_{\lambda_1}(k_1)+\gamma_{\lambda_2}(k_2)\rightarrow f_{\sigma_1}(p_1)+\bar{f}_{\sigma_2}(p_2)\,,\notag
\end{equation}
where $\lambda_i=\pm1$ and $\sigma_i=\pm1$ denote the initial and final state helicities (normalized to one). Then, in the center-of-mass frame the detector weight is given by
\begin{equation}
\label{q5detector}
\hspace{-0.25cm}q_5^{\sigma_1\sigma_2}(\n,\n_p)=\frac{1}{\beta}\left[\sigma_1\d^{(2)}(\n-\n_p)+\sigma_2\d^{(2)}(\n+\n_p)\right]\,,
\end{equation}
where $\p$ is the outgoing fermion momentum and $\beta=\sqrt{1-\frac{4m^{2}}{s}}$. We choose $k_{1,2}=\frac{\sqrt{s}}{2}(1,0,0,\pm1)$ and write the fully polarized amplitude squared as 
\vspace{-0.9cm}
\begin{widetext}
\begin{align}
\label{2gammaAmplitude}
\left| \mathcal M_{\lambda_{1,2}}^{\sigma_{1,2}}(\cos\theta)\right|^2 = & \frac{4e^4(1-\beta^2)} {\left(1-\beta^2\cos^2\theta\right)^2} \Bigg[ \frac{1+\lambda_1\lambda_2}{2} \left( 1+ \frac{\lambda_1+\lambda_2}{2}\, \sigma_1\beta \right)^2 \delta_{\sigma_1,\sigma_2} \nonumber\\ &\hspace{2cm}+ \frac{1-\lambda_1\lambda_2}{2} \beta^2 \Bigg\{\sin^4\theta\delta_{\sigma_1,\sigma_2} + \frac{\sin^2\theta}{(1-\beta^2)}\left( 1+ \frac{\lambda_1-\lambda_2}{2}\, \sigma_1\cos\theta \right)^2 \delta_{\sigma_1,-\sigma_2} \Bigg\} \Bigg]\,,
\end{align}
where $\cos\theta=p_z/|\p|$.\\

\noindent The resulting axial-flux one-point function, depending on the polarization of the initial photons, reads
\begin{align*}
\hspace{0cm}\Sigma_5^{\lambda_{1,2}}=\frac{1}{64s\pi^2\sigma^{\lambda_{1,2}}_{\rm tot}}\sum_{\sigma_{1,2}}\Bigg[\s_1\left| \mathcal M_{\lambda_{1,2}}^{\sigma_{1,2}}(n_z)\right|^2+\s_2\left| \mathcal M_{\lambda_{1,2}}^{\sigma_{1,2}}(-n_z)\right|^2\Bigg]
=\frac{4\alpha^2}{s\sigma^{\lambda_{1,2}}_{\rm tot}}\left[\frac{(\lambda_1+\lambda_2)\beta(1-\beta^2)}{(1-\beta^2n_z^2)^2}+\frac{(\lambda_1-\lambda_2)\beta^2n_z(1-n_z^2)}{(1-\beta^2n_z^2)^2}\right]\,,
\end{align*}
\end{widetext}
where $\sigma^{\lambda_{1,2}}_{\rm tot}$ is kept implicit for compactness and $n_z$ controls the angle with respect to the beam direction.

Taking the fixed-angle massless limit for $|n_z|<1$, the remaining contribution is odd with respect to $n_z$ and gives no net flux under the symmetric integration. Nevertheless, the full flux reads
\begin{align}
&\lim_{m\to 0}\int d^2\n\, \Sigma_5^{\lambda_{1,2}}(\n)=
(\lambda_1+\lambda_2)\,.
\end{align}
Thus, one may see that the massless limit in the same helicity channel is subtle \cite{Dolgov:1971ri}, and the axial-flux one-point function is given by
\begin{align}
&\lim_{m\to0}\Sigma_5^{\lambda\lambda}(\n)=\frac{\lambda}{2\pi}\Big\{\delta(1-n_z)+\delta(1+n_z)\Big\}\,.
\label{anomalousQ5C}
\end{align}

This gives an explicit detector realization of the axial anomaly and the associated IR mechanism. In the classically massless theory, axial-charge conservation would imply a vanishing net outgoing flux. Instead, the anomalous helicity-flip contribution survives angular integration. At the perturbative order considered here, the Ward identity on the fermion--antifermion cut is saturated by the mass term. Its contribution vanishes at every fixed non-collinear angle but survives angular integration in the chiral limit, reproducing the anomalous helicity-flip mechanism underlying the anomaly sum rule.

In more general terms, the axial-flow detector puts the states crossing the cut on-shell. The resulting phase-space integral is then weighted by the eigenvalue of the flow detector on the cut state. It is instructive to compare this detector representation with the dispersive description of the anomalous triangle. Let $p$ and $q$ be the momenta entering the two vector vertices, with $k=p+q$ the momentum flowing out of the axial vertex. The triangle amplitude, $\Gamma^{\mu\alpha\beta}(p,q)$, can be reconstructed dispersively from its discontinuity, which is schematically given by
\begin{equation}
  2\,\operatorname{Im} \Gamma^{\mu\alpha\beta}(p,q) = \sum_X \left[ \mathcal M_{5\rightarrow X}^{\mu} \right]^{*} \mathcal M_{\gamma\gamma\rightarrow X}^{\alpha\beta}(p,q)\,,
\end{equation}
where $\mathcal{M}_{5\rightarrow X}^{\mu}=\langle X|J_5^\mu(0)|0\rangle$, while $\mathcal M_{\gamma\gamma\rightarrow X}^{\alpha\beta}(p,q)$ is the matrix element generated by the two vector currents. Comparing with Eq.~\eqref{Q5generalX}, one sees that the same on-shell channel enters both constructions, though through different matrix elements and with different weights.

Requiring Lorentz covariance, Bose symmetry, and the vector Ward identities, we can further fix the tensor decomposition of the anomalous amplitude \cite{Dolgov:1971ri,Giannotti:2008cv}. Focusing on the on-shell photons $p^2=q^2=0$, one finds that the divergence of the triangle contains only one independent tensor structure
\begin{equation} 
\Gamma^{\mu\alpha\beta}(p,q) = k^\mu F_A(k^2)\, \epsilon^{\alpha\beta\rho\sigma}p_\rho q_\sigma+\Gamma_\perp^{\mu\alpha\beta}\,,
\end{equation}
where the transverse part satisfies $k_\mu\Gamma_\perp^{\mu\alpha\beta}=0$. At leading order, the spectral density of the longitudinal form factor reads
\begin{equation}
\rho_A(s) = \frac{1}{\pi} \operatorname{Im}F_A(s)= \frac{e^2}{\pi^2} \frac{m^2}{s^2} \ln\left( \frac{1+\beta}{1-\beta} \right) \theta(s-4m^2)\,.\notag
\end{equation} 
At every fixed $s>0$, this expression vanishes as $m$ goes to zero, while its integral is fixed
\begin{align}
\int_0^\infty ds\, \rho_A(s) = \frac{e^2}{2\pi^2}\,,
\end{align}
hence $\lim_{m\to0}\rho_A(s) =\frac{e^2}{2\pi^2}\delta(s)$ and the spectral weight therefore collapses onto a massless pseudoscalar channel for $m=0$:
\begin{align}
\Gamma^{\mu\alpha\beta}(p,q) =\frac{e^2}{2\pi^2} \frac{k^\mu}{k^2} \epsilon^{\alpha\beta\rho\sigma}p_\rho q_\sigma+\Gamma_\perp^{\mu\alpha\beta}.
\end{align}
Thus, the anomalous spectral density is the discontinuity of a linear form factor, whereas the axial-flow correlator Eq.~\eqref{anomalousQ5C} is quadratic in the production amplitude and weighted by the axial charge of the cut state. Their singular limits therefore occur in different resolved variables: the spectral density collapses onto zero invariant mass, while the axial-flow distribution at fixed $s$ becomes localized in the two beam-collinear directions. In this sense, the axial-flow correlator provides an angular detector counterpart of the spectral collapse that produces the anomaly pole.

This also holds in the general situation --- the anomaly fixes the integrated axial flux, while the initial state $|\Psi\rangle $ and the
IR dynamics determine how this flux is distributed over angle and energy. For instance, for an isotropic state generated by a pseudoscalar source, rotational invariance removes the angular dependence and the anomalous sum rule fixes the one-point function completely. Another concrete non-perturbative realization occurs in QCD. In the non-singlet axial channel, the perturbative anomaly pole is matched by the pion pole in the chiral limit. The integrated detector sum rule therefore survives confinement, while its angular and energetic content is redistributed from partonic fermion states into pion-containing hadronic channels, suggesting that pion-resolved detector correlations could provide a hadronic probe of the anomaly.

The same construction also extends to the axial anomaly channels, such as the mixed gravitational anomaly \cite{Kimura:1969iwz,Delbourgo:1972xb,Eguchi:1976db,Alvarez-Gaume:1983ihn} given by 
\begin{align}
  \nabla_{\mu}J_{5}^{\mu}
  =2im\,\bar\psi\gamma_{5}\psi+\frac{1}{384\pi^2}\epsilon^{\mu\nu\alpha\beta}R^{\rho}_{~\sigma\mu\nu}R^{\sigma}_{~\rho\alpha\beta}\,,
\end{align}
which also exhibits an analogous mass-independent spectral sum rule \cite{Coriano:2023gxa,Coriano:2025ceu}. Moreover, a gravitational anomaly also arises for the photon helicity current \cite{Dolgov:1987yp,Dolgov:1987xv,Vainshtein:1988ww,Dolgov:1988qx,Dolgov:1988ga,Erdmenger:1999xx}:
\begin{align}
  \nabla_{\mu}K^{\mu}
  =\frac{1}{192\pi^2}\epsilon^{\mu\nu\alpha\beta}R^{\rho}_{~\sigma\mu\nu}R^{\sigma}_{~\rho\alpha\beta}\,,
\end{align}
where $K^{\mu}=\epsilon^{\mu\nu\alpha\beta}A_\nu\pa_\alpha A_\beta$ \footnote{Notice that this anomaly re-appears in different measures of the helicity current, see e.g. \cite{Agullo:2016lkj,Agullo:2018nfv,Galaverni:2020xrq}} and we assume that $\langle \tvec{E}\cdot\tvec{B}\rangle=0$ to isolate the gravitational term. Thus, similar anomalous contributions could be expected in the axial-flux one-point function in the gravitational channel and in the analogous helicity-flux one-point function for chiral cut states of different spin.

\noindent\textbf{Higher-spin helicity detectors:} Having discussed how the axial detector resolves the anomaly-related cut, it is natural to ask whether further information about the anomalous physics can be extracted by changing the detector within a family of operators related to the axial current. For instance, one may consider the energy weighting which has recently become a standard refinement of flow observables, see e.g. \cite{Chen:2020vvp,Riembau:2024tom,Budhraja:2025ulx,Alipour-fard:2025dvp}. A particularly illuminating choice for such a family of generalized axial fluxes can be constructed with higher-spin helicity currents which can be introduced for free fields. Recent interest in polarization transport of photons attracted attention to these objects, known as zilches in that case, see \cite{Chernodub:2018era, Huang:2020kik,Hattori:2020gqh}. They admit a closely analogous realization for fermions \cite{Alexandrov:2020zsj}, and here we will refer to them as zilches collectively. Studying these detectors allows us to test whether the same singular contribution found for the axial flux persists in its higher energy moments. As we show below, this persistence provides detector-level evidence for anomalous contributions to the divergences of the higher-spin currents \footnote{Closely related higher-spin generalizations of the axial anomaly were considered in the literature, see e.g. \cite{Bass:1993ym,Mueller:1997zu}}, although we do not derive their complete operator form here.

Focusing on the fermionic case, we start by considering a tower of parity-odd symmetric higher-spin currents conserved for massless free fermions
\begin{align} Z^{\alpha_1\cdots\alpha_l} = \bar\psi\, \gamma^{\{\alpha_1} i\overleftrightarrow{\partial}^{\alpha_2} \cdots i\overleftrightarrow{\partial}^{\alpha_l\}} \gamma_5\psi\,,
\end{align}
where $l$ is odd, $\overleftrightarrow{\partial}^{\mu} \equiv \frac{1}{2} \left( \overrightarrow{\partial}^{\mu} - \overleftarrow{\partial}^{\mu} \right)$ is the bidirectional derivative, and $A^{\{\alpha_1\cdots\alpha_l\}} \equiv \frac{1}{l!} \sum A^{\Pi(\alpha_1...\alpha_l)}$ with the sum going over all permutations $\Pi$. One may also notice that the axial current is the lowest one in this tower, corresponding to $l=1$. Working in the lab frame, we can readily introduce the generalized zilch-flux operator
\begin{align}
  \mathcal{Z}_{l}(\n)=\lim_{R\rightarrow\infty}
  R^{2}\int_{0}^{\infty}dt\,n_{i}Z^{i0\cdots 0}(t,R\n)\,.
\end{align}
On a one-particle state it measures an energy-weighted helicity flux
\begin{align}
  \mathcal{Z}_{l}(\n)|\p,\sigma\rangle=\sigma E^{l-1}_p \left(\frac{1+(l-1)\beta^2_p}{l \beta_p}\right)\,\delta^{(2)}(\n-\n_p)|\p,\sigma\rangle\,.\notag
\end{align}
Turning to the same leading perturbative process, we notice that the analogue of our detector weight on the two-fermion state in the center-of-mass frame of the two initial state photons in Eq.~\eqref{q5detector} becomes
\begin{align}
  &\hspace{-0.2cm}q_{Z,l}^{\sigma_1\sigma_2}(\n,\n_p)=\frac{1+(l-1)\beta^2}{l}\hspace{-0.1cm}\left(\frac{\sqrt{s}}{2}\right)^{l-1}\hspace{-0.2cm}q_5^{\sigma_1\sigma_2}(\n,\n_p)\,.
\end{align}
Introducing the tower of zilch-flux detectors $\Sigma_{Z,l}(\n)=\langle\Psi|\mathcal{Z}_l(\n)|\Psi\rangle$ and focusing on the particular equal-energy two-body state, we find that all members of the tower are proportional to the axial-flux one-point function
\begin{align}
&\Sigma_{Z,l}^{\lambda_{1,2}}(\n)=\left(\frac{1+(l-1)\beta^2}{l}\right)\left(\frac{\sqrt{s}}{2}\right)^{l-1}\Sigma_{5}^{\lambda_{1,2}}(\n)\,.
\end{align}
In the massless limit, we find the same singular behavior in the same helicity channel as in the case of the axial-flux one-point function
\begin{align}
&\hspace{-0.3cm}\lim_{m\to0}\Sigma_{Z,l}^{\lambda\lambda}=\frac{\lambda}{2\pi}\left(\frac{\sqrt{s}}{2}\right)^{l-1}\Big\{\delta(1-n_z)+\delta(1+n_z)\Big\}\,,
\end{align}
and the corresponding sum rule reads
\begin{align}
&\lim_{m\to0}\int d^2\n\,\Sigma_{Z,l}^{\lambda_{1,2}}(\n)=\left(\frac{\sqrt{s}}{2}\right)^{l-1}(\lambda_1+\lambda_2)\,,
\end{align}
showing that the anomalous helicity-flip contribution persists throughout the energy-weighted tower.

The higher-spin detectors therefore measure the same singular contribution as the axial detector, but weighted by energy moments. While for the simple two-to-two process considered above the energy moments enter the zilch-flux one-point functions trivially, it is not the case in general, and the genuinely new information appears for states with a non-trivial energy distribution. This tower is closely related to earlier higher-spin and point-split generalizations of the axial anomaly, see e.g. \cite{Bass:1993ym,Mueller:1997zu}. Thus, while the general zilch vortical effects suggest a possible anomalous origin for the higher-spin tower \cite{Huang:2020kik,Alexandrov:2020zsj}, the detector construction gives a direct fixed-state manifestation of the same singular chiral-limit mechanism through which the axial anomaly is realized.

To make the detector realization of the axial anomaly more direct and connect it explicitly to the anomalous divergence, we can replace one of the incoming photons by an external magnetic background $\tvec{B}$. A propagating photon or an external electric-field perturbation can then generate axial charge. For an initial state whose evolution produces a single fermion–antifermion pair, we can readily determine the corresponding zilch-flux one-point function. In the strong-field, low-energy regime, the Higher Landau Levels are kinematically inaccessible and the produced fermions occupy the Lowest Landau Level (LLL), where the transverse motion is frozen and the dynamics is effectively reduced to $1+1$ dimensions along $\tvec{B}$. This is the standard strong-field realization of the axial anomaly: the dynamics is projected onto the LLL, while the electric field drives spectral flow \cite{Nielsen:1983rb}, reproducing the two-dimensional axial anomaly. In the pair center-of-mass frame, the fermion and antifermion are emitted in the two opposite directions along $\tvec{B}$ with equal energies $E$. Assuming that no axial charge remains in the interaction region at late times, we find at $m=0$
\begin{align}
&\langle \Psi|\mathcal{Z}_{l}(\n)|\Psi\rangle\Big|_B
=
\frac{E^{l-1}}{2}
\left\langle \Psi\left|
\frac{2\alpha}{\pi}
\int d^4x\,\tvec{E}\cdot\tvec{B}\right|\Psi
\right\rangle
\notag\\
&\hspace{1cm}\times
\left[
\delta^{(2)}(\n-\n_B)
+
\delta^{(2)}(\n+\n_B)
\right]\,,
\end{align}
where $\n_B$ is the unit vector along the magnetic field, $|\Psi\rangle$ is the specific in-state with the background field, and $\left\langle
\frac{2\alpha}{\pi}
\int d^4x\,\tvec{E}\cdot\tvec{B}
\right\rangle$ is the net axial charge generated by the anomaly in the event. Thus, in this LLL state every
higher-spin detector moment is sourced directly by the anomalous divergence, with the characteristic energy weight $E^{l-1}$. This result establishes a further detector-level connection between the axial anomaly and the divergences of the higher-spin zilch currents.

Turning to the mixed gravitational anomaly, we start with the two-graviton scattering into a fermion--antifermion state which closely follows the structure of the two-photon scattering considered above. The integrated (non-normalized) flux carries additional powers of the fermion mass in the potentially anomalous same-helicity channel. However, these are precisely compensated by the scaling of the total cross section with $m$. Thus, we arrive at the same anomalous helicity-flip mechanism: while $\lim_{m\rightarrow 0}
\Sigma^{\lambda\lambda}_{Z,l}(\n)=0$ for $|n_z|<1$, the net normalized flux is non-zero $\lim_{m\rightarrow 0}\int_{\n}\Sigma^{\lambda\lambda}_{Z,l}(\n)\neq 0$. The normalized angular distributions lose uniform integrability in the massless limit, exhibiting the same singular rearrangement of spectral weight that underlies the mixed gravitational anomaly pole:
\begin{align}
&\hspace{-0.3cm}\lim_{m\to0}\Sigma_{Z,l}^{\lambda\lambda}=\frac{\lambda}{2\pi}\left(\frac{\sqrt{s}}{2}\right)^{l-1}\Big\{\delta(1-n_z)+\delta(1+n_z)\Big\}\,,
\end{align}
where $\lambda=\pm1$ labels the physical graviton helicities $\pm2$. In this sense, the zilch-flux one-point functions provide a detector-level counterpart of the mixed gravitational anomaly and show that the singular contribution persists.

Extending the same detector construction to the corresponding helicity currents carried by other massless particles suggests a two-parameter family of anomalous detector sum rules, labeled by the helicity of the particles saturating the cut, $h=\frac{1}{2}, 1, \frac{3}{2}, ...$, and by the spin $l$ of the detected zilch current. We leave the derivation of the corresponding operator-level sum rules and explicit anomalous divergences for future work.

\noindent\textbf{Conclusions:} In this work we have studied how quantum anomalies are realized in detector observables constructed from the corresponding anomalous currents. For the axial anomaly, an axial-flux detector $\mathcal{Q}_5$ probes the fermionic cut of the triangle diagram and resolves it as an angular distribution. In the simplest example of polarized two-photon scattering, the mass-suppressed helicity-flip contribution vanishes at every fixed non-collinear angle in the massless limit, while its angular integral remains finite. Thus, although axial-charge conservation in the classically massless theory would imply a vanishing net outgoing flux, the helicity-flip contribution survives through a discontinuous transfer of spectral weight to the beam-collinear directions. Since the underlying sum rule in Eq.~\eqref{Q5sumrule} is an operator identity, the same relation also applies inside higher-point flow correlations.

Replacing the axial detector by higher-spin helicity detectors, we have shown that the same singular helicity-flip contribution persists throughout the energy-weighted tower of zilches. In the strong-magnetic-field realization, the dynamics reduces to the LLL and every zilch moment is directly sourced by the integrated anomalous divergence. This provides a direct detector-level manifestation of the axial anomaly in the higher-spin fluxes.

We have also extended the analysis to the mixed axial-gravitational anomaly, where the same singular localization mechanism and finite normalized zilch-flux sum rules persist. Together with the gravitational anomaly of photon helicity and the known hierarchy of chiral and zilch vortical effects, this result suggests a two-parameter family of anomalous detector sum rules labeled by the helicity of the states saturating the cut and by the spin of the detected current. 

Finally, 
the discussion developed here naturally extends to phenomenological contexts. 
Matching the perturbative anomaly pole onto the pion channel suggests that pion-resolved detector correlations may provide a hadronic probe of the anomaly. In $e^+e^-$ collisions, crossed electroweak assignments can move the axial insertion to the source and probe the same anomalous amplitude through $\mathcal{Q}$ detectors, accessible from recently revived archival data.

\noindent\textbf{Acknowledgments:} We would like to thank P. Morales, E. Mottola, N. Poovuttikul, V. I. Zakharov, and H.-X. Zhu for useful discussions. The work of AVS is supported by Funda\symbol{"00E7}\symbol{"00E3}o para a Ci\symbol{"00EA}ncia e a Tecnologia (FCT) under contract 
2023.15319.PEX (https://doi.org/10.54499/2023.15319.PEX) and by the Basque Government through grant IT1628-22. AVS would also like to acknowledge support from Ikerbasque, Basque Foundation for Science.

\bibliographystyle{bibstyle.bst}
\bibliography{anomalies.bib}

@book{Bertlbook,
  title = {Anomalies in Quantum Field Theory},
  author = {Bertlmann, Reinhold A.},
@series = {International Series of Monographs on Physics},
  year = {1996}, 
  publisher = {Oxford Univ. Press},
}

@article{Giannotti:2008cv,
    author = "Giannotti, Maurizio and Mottola, Emil",
    title = "{The Trace Anomaly and Massless Scalar Degrees of Freedom in Gravity}",
    eprint = "0812.0351",
    archivePrefix = "arXiv",
    primaryClass = "hep-th",
    reportNumber = "LA-UR-08-6329",
    doi = "10.1103/PhysRevD.79.045014",
    journal = "Phys. Rev. D",
    volume = "79",
    pages = "045014",
    year = "2009"
}

@article{Mottola:2019nui,
    author = "Mottola, Emil and Sadofyev, Andrey V.",
    title = "{Chiral Waves on the Fermi-Dirac Sea: Quantum Superfluidity and the Axial Anomaly}",
    eprint = "1909.01974",
    archivePrefix = "arXiv",
    primaryClass = "hep-th",
    reportNumber = "LA-UR-19-27117",
    doi = "10.1016/j.nuclphysb.2021.115385",
    journal = "Nucl. Phys. B",
    volume = "966",
    pages = "115385",
    year = "2021"
}

@article{Mottola:2023emy,
    author = "Mottola, Emil and Sadofyev, Andrey V. and Stergiou, Andreas",
    title = "{Axions and superfluidity in Weyl semimetals}",
    eprint = "2310.08629",
    archivePrefix = "arXiv",
    primaryClass = "hep-th",
    doi = "10.1103/PhysRevB.109.134512",
    journal = "Phys. Rev. B",
    volume = "109",
    number = "13",
    pages = "134512",
    year = "2024"
}

@article{Huang:2015oca,
      author         = "Huang, Xu-Guang",
      title          = "{Electromagnetic fields and anomalous transports in
                        heavy-ion collisions --- A pedagogical review}",
      journal        = "Rept. Prog. Phys.",
      volume         = "79",
      year           = "2016",
      number         = "7",
      pages          = "076302",
      doi            = "10.1088/0034-4885/79/7/076302",
      eprint         = "1509.04073",
      archivePrefix  = "arXiv",
      primaryClass   = "nucl-th"
}

@article{Moult:2025nhu,
    author = "Moult, Ian and Zhu, Hua Xing",
    title = "{Energy Correlators: A Journey From Theory to Experiment}",
    eprint = "2506.09119",
    archivePrefix = "arXiv",
    primaryClass = "hep-ph",
    month = "6",
    year = "2025"
}

@article{Adler:1969gk,
      author         = "Adler, Stephen L.",
      title          = "{Axial vector vertex in spinor electrodynamics}",
      journal        = "Phys. Rev.",
      volume         = "177",
      year           = "1969",
      pages          = "2426-2438",
      doi            = "10.1103/PhysRev.177.2426"
}

@article{Collins:1976yq,
    author = "Collins, John C. and Duncan, Anthony and Joglekar, Satish D.",
    title = "{Trace and Dilatation Anomalies in Gauge Theories}",
    reportNumber = "COO-2220-88",
    doi = "10.1103/PhysRevD.16.438",
    journal = "Phys. Rev. D",
    volume = "16",
    pages = "438--449",
    year = "1977"
}

@article{Adler:1976zt,
    author = "Adler, Stephen L. and Collins, John C. and Duncan, Anthony",
    title = "{Energy-Momentum-Tensor Trace Anomaly in Spin 1/2 Quantum Electrodynamics}",
    reportNumber = "COO-2220-77-REV, COO-2220-77",
    doi = "10.1103/PhysRevD.15.1712",
    journal = "Phys. Rev. D",
    volume = "15",
    pages = "1712",
    year = "1977"
}

@article{lipkin,
      author         = "Lipkin, Daniel M.",
      title          = "{Existence of a New Conservation Law in Electromagnetic
      Theory}",
      journal        = "Journal of Mathematical Physics",
      volume         = "5",
      year           = "1964",
      pages          = "696",
      doi            = "doi.org/10.1063/1.1704165",
      SLACcitation   = "%%CITATION = ARXIV:1805.08779;%%"
}

@article{kibble,
      author         = "Kibble, T. W. B.",
      title          = "{Conservation Laws for Free Fields}",
      journal        = "Journal of Mathematical Physics",
      volume         = "6",
      year           = "1965",
      pages          = "1022",
      doi            = "doi.org/10.1063/1.1704363",
      SLACcitation   = "%%CITATION = ARXIV:1805.08779;%%"
}

@article{Huang:2020kik,
    author = "Huang, Xu-Guang and Mitkin, Pavel and Sadofyev, Andrey V. and Speranza, Enrico",
    title = "{Zilch Vortical Effect, Berry Phase, and Kinetic Theory}",
    eprint = "2006.03591",
    archivePrefix = "arXiv",
    primaryClass = "hep-th",
    doi = "10.1007/JHEP10(2020)117",
    journal = "JHEP",
    volume = "10",
    pages = "117",
    year = "2020"
}

@article{Huang:2018aly,
    author = "Huang, Xu-Guang and Sadofyev, Andrey V.",
    title = "{Chiral Vortical Effect For An Arbitrary Spin}",
    eprint = "1805.08779",
    archivePrefix = "arXiv",
    primaryClass = "hep-th",
    doi = "10.1007/JHEP03(2019)084",
    journal = "JHEP",
    volume = "03",
    pages = "084",
    year = "2019"
}

@article{Avkhadiev:2017fxj,
    author = "Avkhadiev, Artur and Sadofyev, Andrey V.",
    title = "{Chiral Vortical Effect for Bosons}",
    eprint = "1702.07340",
    archivePrefix = "arXiv",
    primaryClass = "hep-th",
    reportNumber = "MIT-CTP-4884",
    doi = "10.1103/PhysRevD.96.045015",
    journal = "Phys. Rev. D",
    volume = "96",
    number = "4",
    pages = "045015",
    year = "2017"
}

@article{Hattori:2020gqh,
    author = "Hattori, Koichi and Hidaka, Yoshimasa and Yamamoto, Naoki and Yang, Di-Lun",
    title = "{Wigner functions and quantum kinetic theory of polarized photons}",
    eprint = "2010.13368",
    archivePrefix = "arXiv",
    primaryClass = "hep-ph",
    reportNumber = "YITP-20-129, KEK-TH-2262, J-PARC-TH-0228, RIKEN-iTHEMS-Report-20",
    doi = "10.1007/JHEP02(2021)001",
    journal = "JHEP",
    volume = "02",
    pages = "001",
    year = "2021"
}

@article{Alexandrov:2020zsj,
    author = "Alexandrov, Artem and Mitkin, Pavel",
    title = "{Zilch Vortical Effect for Fermions}",
    eprint = "2011.09429",
    archivePrefix = "arXiv",
    primaryClass = "hep-th",
    doi = "10.1007/JHEP05(2021)070",
    journal = "JHEP",
    volume = "05",
    pages = "070",
    year = "2021"
}

@article{Yamamoto:2017uul,
    author = "Yamamoto, Naoki",
    title = "{Photonic chiral vortical effect}",
    eprint = "1702.08886",
    archivePrefix = "arXiv",
    primaryClass = "hep-th",
    doi = "10.1103/PhysRevD.96.051902",
    journal = "Phys. Rev. D",
    volume = "96",
    number = "5",
    pages = "051902",
    year = "2017"
}

@article{Chernodub:2018era,
    author = "Chernodub, M. N. and Cortijo, Alberto and Landsteiner, Karl",
    title = "{Zilch vortical effect}",
    eprint = "1807.10705",
    archivePrefix = "arXiv",
    primaryClass = "hep-th",
    reportNumber = "IFT-UAM/CSIC-18-085",
    doi = "10.1103/PhysRevD.98.065016",
    journal = "Phys. Rev. D",
    volume = "98",
    number = "6",
    pages = "065016",
    year = "2018"
}

@article{Hayata:2017tbr,
    author = "Hayata, Tomoya",
    title = "{Chiral magnetic effect of light}",
    eprint = "1705.09926",
    archivePrefix = "arXiv",
    primaryClass = "physics.optics",
    doi = "10.1103/PhysRevB.97.205102",
    journal = "Phys. Rev. B",
    volume = "97",
    number = "20",
    pages = "205102",
    year = "2018"
}

@article{Dwivedi:2016kkj,
    author = "Dwivedi, Vatsal and Stone, Michael",
    title = "{Chiral kinetic theory and anomalous hydrodynamics in even spacetime dimensions}",
    eprint = "1606.04945",
    archivePrefix = "arXiv",
    primaryClass = "hep-th",
    doi = "10.1088/1751-8121/aa626",
    journal = "J. Phys. A",
    volume = "50",
    number = "15",
    pages = "155202",
    year = "2017"
}

@article{Loganayagam:2012pz,
    author = "Loganayagam, R. and Surowka, Piotr",
    title = "{Anomaly/Transport in an Ideal Weyl gas}",
    eprint = "1201.2812",
    archivePrefix = "arXiv",
    primaryClass = "hep-th",
    doi = "10.1007/JHEP04(2012)097",
    journal = "JHEP",
    volume = "04",
    pages = "097",
    year = "2012"
}

@article{Banerjee:2012cr,
    author = "Banerjee, Nabamita and Dutta, Suvankar and Jain, Sachin and Loganayagam, R. and Sharma, Tarun",
    title = "{Constraints on Anomalous Fluid in Arbitrary Dimensions}",
    eprint = "1206.6499",
    archivePrefix = "arXiv",
    primaryClass = "hep-th",
    doi = "10.1007/JHEP03(2013)048",
    journal = "JHEP",
    volume = "03",
    pages = "048",
    year = "2013"
}

@article{Stone:2018zel,
    author = "Stone, Michael and Kim, Jiyoung",
    title = "{Mixed Anomalies: Chiral Vortical Effect and the Sommerfeld Expansion}",
    eprint = "1804.08668",
    archivePrefix = "arXiv",
    primaryClass = "cond-mat.mes-hall",
    doi = "10.1103/PhysRevD.98.025012",
    journal = "Phys. Rev. D",
    volume = "98",
    number = "2",
    pages = "025012",
    year = "2018"
}

@article{Nielsen:1977sy,
    author = "Nielsen, N. K.",
    title = "{The Energy Momentum Tensor in a Nonabelian Quark Gluon Theory}",
    doi = "10.1016/0550-3213(77)90040-2",
    journal = "Nucl. Phys. B",
    volume = "120",
    pages = "212--220",
    year = "1977"
}

@article{Duff:1977ay,
    author = "Duff, M. J.",
    title = "{Observations on Conformal Anomalies}",
    reportNumber = "Print-77-0361 (QUEEN MARY COLL.)",
    doi = "10.1016/0550-3213(77)90410-2",
    journal = "Nucl. Phys. B",
    volume = "125",
    pages = "334--348",
    year = "1977"
}

@article{Eguchi:1976db,
    author = "Eguchi, Tohru and Freund, Peter G. O.",
    title = "{Quantum Gravity and World Topology}",
    reportNumber = "EFI 76/52-CHICAGO",
    doi = "10.1103/PhysRevLett.37.1251",
    journal = "Phys. Rev. Lett.",
    volume = "37",
    pages = "1251",
    year = "1976"
}

@article{Galaverni:2020xrq,
    author = "Galaverni, Matteo and Gabriele, S. J., Gionti",
    title = "{Photon helicity and quantum anomalies in curved spacetimes}",
    eprint = "2012.02583",
    archivePrefix = "arXiv",
    primaryClass = "gr-qc",
    doi = "10.1007/s10714-021-02817-z",
    journal = "Gen. Rel. Grav.",
    volume = "53",
    number = "4",
    pages = "46",
    year = "2021"
}

@article{Armillis:2009im,
    author = "Armillis, Roberta and Coriano, Claudio and Delle Rose, Luigi",
    title = "{Anomaly Poles as Common Signatures of Chiral and Conformal Anomalies}",
    eprint = "0909.4522",
    archivePrefix = "arXiv",
    primaryClass = "hep-ph",
    doi = "10.1016/j.physletb.2009.11.013",
    journal = "Phys. Lett. B",
    volume = "682",
    pages = "322--327",
    year = "2009"
}

@article{Coriano:2023gxa,
    author = "Corian{\`o}, Claudio and Lionetti, Stefano and Maglio, Matteo Maria",
    title = "{Parity-violating CFT and the gravitational chiral anomaly}",
    eprint = "2309.05374",
    archivePrefix = "arXiv",
    primaryClass = "hep-th",
    doi = "10.1103/PhysRevD.109.045004",
    journal = "Phys. Rev. D",
    volume = "109",
    number = "4",
    pages = "045004",
    year = "2024"
}

@article{Mueller:1997zu,
    author = "Mueller, Dieter and Teryaev, O. V.",
    title = "{Nonlocal generalization of the axial anomaly and x dependence of the anomalous gluon contribution}",
    eprint = "hep-ph/9701413",
    archivePrefix = "arXiv",
    reportNumber = "CERN-TH-97-006, CERN-TH-97-06, CERN-TH-97-6",
    doi = "10.1103/PhysRevD.56.2607",
    journal = "Phys. Rev. D",
    volume = "56",
    pages = "2607--2613",
    year = "1997"
}

@article{Bass:1993ym,
    author = "Bass, S. D.",
    title = "{Gauge symmetry and the EMC spin effect}",
    eprint = "hep-ph/9305323",
    archivePrefix = "arXiv",
    reportNumber = "CAVENDISH-HEP-93-2",
    doi = "10.1007/BF01474632",
    journal = "Z. Phys. C",
    volume = "60",
    pages = "343--348",
    year = "1993"
}

@article{Dolgov:1987xv,
    author = "Dolgov, A. D. and Khriplovich, I. B. and Zakharov, Valentin I.",
    title = "{MACROSCOPIC MANIFESTATIONS OF THE CHIRAL ANOMALY IN GRAVITATIONAL FIELD}",
    reportNumber = "IYF-87-56",
    doi = "10.1016/0550-3213(88)90460-9",
    journal = "Sov. Phys. JETP",
    volume = "67",
    pages = "237",
    year = "1988"
}

@article{Dolgov:1988ga,
    author = "Dolgov, A. D. and Khriplovich, I. B. and Vainshtein, A. I. and Zakharov, Valentin I.",
    title = "{VANISHING OF THE CHIRAL ANOMALY FOR ANTISYMMETRIC TENSOR FIELD}",
    reportNumber = "IYF-88-51",
    doi = "10.1016/0550-3213(89)90513-0",
    journal = "Nucl. Phys. B",
    volume = "313",
    pages = "73--79",
    year = "1989"
}

@article{Agullo:2018nfv,
    author = "Agullo, Ivan and del Rio, Adrian and Navarro-Salas, Jose",
    title = "{Classical and quantum aspects of electric-magnetic duality rotations in curved spacetimes}",
    eprint = "1810.08085",
    archivePrefix = "arXiv",
    primaryClass = "gr-qc",
    doi = "10.1103/PhysRevD.98.125001",
    journal = "Phys. Rev. D",
    volume = "98",
    number = "12",
    pages = "125001",
    year = "2018"
}

@article{Agullo:2016lkj,
    author = "Agullo, I. and del Rio, A. and Navarro-Salas, J.",
    title = "{Electromagnetic duality anomaly in curved spacetimes}",
    eprint = "1607.08879",
    archivePrefix = "arXiv",
    primaryClass = "gr-qc",
    doi = "10.1103/PhysRevLett.118.111301",
    journal = "Phys. Rev. Lett.",
    volume = "118",
    number = "11",
    pages = "111301",
    year = "2017"
}

@article{Dolgov:1988qx,
    author = "Dolgov, A. D. and Khriplovich, I. B. and Vainshtein, A. I. and Zakharov, Valentin I.",
    title = "{Photonic Chiral Current and Its Anomaly in a Gravitational Field}",
    reportNumber = "IYF-88-37",
    doi = "10.1016/0550-3213(89)90451-3",
    journal = "Nucl. Phys. B",
    volume = "315",
    pages = "138--152",
    year = "1989"
}

@article{Vainshtein:1988ww,
    author = "Vainshtein, A. I. and Dolgov, A. D. and Zakharov, Valentin I. and Khriplovich, I. B.",
    title = "{CHIRAL PHOTON CURRENT AND ITS ANOMALY IN A GRAVITATIONAL FIELD}",
    journal = "Sov. Phys. JETP",
    volume = "67",
    pages = "1326--1332",
    year = "1988"
}

@article{Dolgov:1987yp,
    author = "Dolgov, A. D. and Khriplovich, I. B. and Zakharov, Valentin I.",
    title = "{Chiral Boson Anomaly in a Gravitational Field}",
    journal = "JETP Lett.",
    volume = "45",
    pages = "651--653",
    year = "1987"
}

@article{Alvarez-Gaume:1983ihn,
    author = "Alvarez-Gaume, Luis and Witten, Edward",
    editor = "Salam, A. and Sezgin, E.",
    title = "{Gravitational Anomalies}",
    reportNumber = "HUTP-83/A039",
    doi = "10.1016/0550-3213(84)90066-X",
    journal = "Nucl. Phys. B",
    volume = "234",
    pages = "269",
    year = "1984"
}

@article{Delbourgo:1972xb,
    author = "Delbourgo, Robert and Salam, Abdus",
    title = "{The gravitational correction to pcac}",
    reportNumber = "ICTP-71-24",
    doi = "10.1016/0370-2693(72)90825-8",
    journal = "Phys. Lett. B",
    volume = "40",
    pages = "381--382",
    year = "1972"
}

@article{Prokhorov:2020npf,
    author = "Prokhorov, G. Yu and Teryaev, O. V. and Zakharov, V. I.",
    title = "{Chiral vortical effect for vector fields}",
    eprint = "2009.11402",
    archivePrefix = "arXiv",
    primaryClass = "hep-th",
    doi = "10.1103/PhysRevD.103.085003",
    journal = "Phys. Rev. D",
    volume = "103",
    number = "8",
    pages = "085003",
    year = "2021"
}

@article{Kamada:2022nyt,
    author = "Kamada, Kohei and Yamamoto, Naoki and Yang, Di-Lun",
    title = "{Chiral effects in astrophysics and cosmology}",
    eprint = "2207.09184",
    archivePrefix = "arXiv",
    primaryClass = "astro-ph.CO",
    reportNumber = "RESCEU-12/22",
    doi = "10.1016/j.ppnp.2022.104016",
    journal = "Prog. Part. Nucl. Phys.",
    volume = "129",
    pages = "104016",
    year = "2023"
}

@article{Crewther:1972kn,
    author = "Crewther, R. J.",
    title = "{Nonperturbative evaluation of the anomalies in low-energy theorems}",
    doi = "10.1103/PhysRevLett.28.1421",
    journal = "Phys. Rev. Lett.",
    volume = "28",
    pages = "1421",
    year = "1972"
}

@article{Bell:1969ts,
      author         = "Bell, J. S. and Jackiw, R.",
      title          = "{A PCAC puzzle: pi0 $\to$ gamma gamma in the sigma model}",
      journal        = "Nuovo Cimento",
      volume         = "A60",
      year           = "1969",
      pages          = "47-61",
      doi            = "10.1007/BF02823296"
}

@article{Dolgov:1971ri,
      author         = "Dolgov, A. D. and Zakharov, V. I.",
      title          = "{On Conservation of the axial current in massless
                        electrodynamics}",
      journal        = "Nucl. Phys.",
      volume         = "B27",
      year           = "1971",
      pages          = "525-540",
      doi            = "10.1016/0550-3213(71)90264-1"
}

@article{Kharzeev:2015znc,
    author = "Kharzeev, D. E. and Liao, J. and Voloshin, S. A. and Wang, G.",
    title = "{Chiral magnetic and vortical effects in high-energy nuclear collisions{\textemdash}A status report}",
    eprint = "1511.04050",
    archivePrefix = "arXiv",
    primaryClass = "hep-ph",
    doi = "10.1016/j.ppnp.2016.01.001",
    journal = "Prog. Part. Nucl. Phys.",
    volume = "88",
    pages = "1--28",
    year = "2016"
}

@article{Hartman:2023qdn,
	archiveprefix = {arXiv},
	author = {Hartman, Thomas and Mathys, Gr\'egoire},
	doi = {10.1007/JHEP12(2023)139},
	eprint = {2309.14409},
	journal = {JHEP},
	pages = {139},
	primaryclass = {hep-th},
	title = {{Averaged null energy and the renormalization group}},
	volume = {12},
	year = {2023}}

@article{Ore:1979ry,
	author = {Ore, Jr., F. R. and Sterman, George F.},
	doi = {10.1016/0550-3213(80)90308-9},
	journal = {Nucl. Phys. B},
	pages = {93--118},
	reportnumber = {Print-79-0892 (IAS,PRINCETON)},
	title = {{AN OPERATOR APPROACH TO WEIGHTED CROSS-SECTIONS}},
	volume = {165},
	year = {1980}}

@article{Chicherin:2020azt,
	archiveprefix = {arXiv},
	author = {Chicherin, D. and Henn, J. M. and Sokatchev, E. and Yan, K.},
	doi = {10.1007/JHEP02(2021)053},
	eprint = {2001.10806},
	journal = {JHEP},
	pages = {053},
	primaryclass = {hep-th},
	reportnumber = {LAPTH-003/20, MPP-2020-8},
	title = {{From correlation functions to event shapes in QCD}},
	volume = {02},
	year = {2021}}

@article{Basham:1979gh,
	author = {Basham, C. Louis and Brown, Lowell S. and Ellis, Stephen D. and Love, Sherwin T.},
	doi = {10.1016/0370-2693(79)90601-4},
	journal = {Phys. Lett. B},
	pages = {297--299},
	reportnumber = {RLO-1388-786},
	title = {{Energy Correlations in Perturbative Quantum Chromodynamics: A Conjecture for All Orders}},
	volume = {85},
	year = {1979}}

@article{Basham:1977iq,
	author = {Basham, C. Louis and Brown, Lowell S. and Ellis, S. D. and Love, S. T.},
	doi = {10.1103/PhysRevD.17.2298},
	journal = {Phys. Rev. D},
	pages = {2298},
	reportnumber = {RLO-1388-746},
	title = {{Electron - Positron Annihilation Energy Pattern in Quantum Chromodynamics: Asymptotically Free Perturbation Theory}},
	volume = {17},
	year = {1978}}

@article{Sveshnikov:1995vi,
	archiveprefix = {arXiv},
	author = {Sveshnikov, N.A. and Tkachov, F.V.},
	doi = {10.1016/0370-2693(96)00558-8},
	editor = {Levchenko, B.B. and Savrin, V.I.},
	eprint = {hep-ph/9512370},
	journal = {Phys. Lett. B},
	pages = {403--408},
	title = {{Jets and quantum field theory}},
	volume = {382},
	year = {1996}}

@article{Tkachov:1995kk,
	archiveprefix = {arXiv},
	author = {Tkachov, Fyodor V.},
	doi = {10.1142/S0217751X97002899},
	eprint = {hep-ph/9601308},
	journal = {Int. J. Mod. Phys. A},
	pages = {5411--5529},
	reportnumber = {FERMILAB-PUB-95-191-T-REV, FERMILAB-PUB-95-191-T},
	title = {{Measuring multi - jet structure of hadronic energy flow or What is a jet?}},
	volume = {12},
	year = {1997}}

@article{Korchemsky:1999kt,
	archiveprefix = {arXiv},
	author = {Korchemsky, Gregory P. and Sterman, George F.},
	doi = {10.1016/S0550-3213(99)00308-9},
	eprint = {hep-ph/9902341},
	journal = {Nucl. Phys. B},
	pages = {335--351},
	reportnumber = {ITP-SB-98-73, LPT-ORSAY-98-80},
	title = {{Power corrections to event shapes and factorization}},
	volume = {555},
	year = {1999}}

@article{Hofman:2008ar,
	archiveprefix = {arXiv},
	author = {Hofman, Diego M. and Maldacena, Juan},
	doi = {10.1088/1126-6708/2008/05/012},
	eprint = {0803.1467},
	journal = {JHEP},
	pages = {012},
	primaryclass = {hep-th},
	title = {{Conformal collider physics: Energy and charge correlations}},
	volume = {05},
	year = {2008}}

@article{Belitsky:2013bja,
	archiveprefix = {arXiv},
	author = {Belitsky, A.V. and Hohenegger, S. and Korchemsky, G.P. and Sokatchev, E. and Zhiboedov, A.},
	doi = {10.1016/j.nuclphysb.2014.04.019},
	eprint = {1309.1424},
	journal = {Nucl. Phys. B},
	pages = {206--256},
	primaryclass = {hep-th},
	reportnumber = {CERN-PH-TH-2013-212},
	title = {{Event shapes in $\mathcal{N} = 4$ super-Yang-Mills theory}},
	volume = {884},
	year = {2014}}

@article{Basham:1978bw,
	author = {Basham, C.Louis and Brown, Lowell S. and Ellis, Stephen D. and Love, Sherwin T.},
	doi = {10.1103/PhysRevLett.41.1585},
	journal = {Phys. Rev. Lett.},
	pages = {1585},
	reportnumber = {RLO-1388-759},
	title = {{Energy Correlations in electron - Positron Annihilation: Testing QCD}},
	volume = {41},
	year = {1978}}

@article{Basham:1978zq,
	author = {Basham, C.L. and Brown, L.S. and Ellis, S.D. and Love, S.T.},
	doi = {10.1103/PhysRevD.19.2018},
	journal = {Phys. Rev. D},
	pages = {2018},
	reportnumber = {RLO-1388-761},
	title = {{Energy Correlations in electron-Positron Annihilation in Quantum Chromodynamics: Asymptotically Free Perturbation Theory}},
	volume = {19},
	year = {1979}}

@article{Chen:2020vvp,
	archiveprefix = {arXiv},
	author = {Chen, Hao and Moult, Ian and Zhang, XiaoYuan and Zhu, Hua Xing},
	doi = {10.1103/PhysRevD.102.054012},
	eprint = {2004.11381},
	journal = {Phys. Rev. D},
	number = {5},
	pages = {054012},
	primaryclass = {hep-ph},
	title = {{Rethinking jets with energy correlators: Tracks, resummation, and analytic continuation}},
	volume = {102},
	year = {2020}}

@article{Belitsky:2014zha,
	archiveprefix = {arXiv},
	author = {Belitsky, A.V. and Hohenegger, S. and Korchemsky, G.P. and Sokatchev, E.},
	doi = {10.1016/j.nuclphysb.2016.01.008},
	eprint = {1409.2502},
	journal = {Nucl. Phys. B},
	pages = {176--215},
	primaryclass = {hep-th},
	reportnumber = {CERN-PH-TH-2014-174, IPHT-T14-122, LAPTH-109-14},
	title = {{N=4 superconformal Ward identities for correlation functions}},
	volume = {904},
	year = {2016}}

@article{Hartman:2016lgu,
    author = "Hartman, Thomas and Kundu, Sandipan and Tajdini, Amirhossein",
    title = "{Averaged Null Energy Condition from Causality}",
    eprint = "1610.05308",
    archivePrefix = "arXiv",
    primaryClass = "hep-th",
    doi = "10.1007/JHEP07(2017)066",
    journal = "JHEP",
    volume = "07",
    pages = "066",
    year = "2017"
}

@article{Electron-PositronAlliance:2025fhk,
    author = "Bossi, Hannah and others",
    collaboration = "Electron-Positron Alliance",
    title = "{Energy Correlators from Partons to Hadrons: Unveiling the Dynamics of the Strong Interactions with Archival ALEPH Data}",
    eprint = "2511.00149",
    archivePrefix = "arXiv",
    primaryClass = "hep-ph",
    reportNumber = "MITP-25-057, MITHIG-MOD-24-001",
    month = "10",
    year = "2025"
}

@article{Budhraja:2025ulx,
    author = "Budhraja, Ankita and Singh, Balbeer",
    title = "{Exploiting {\ensuremath{\nu}}-dependence of projected energy correlators in HICs}",
    eprint = "2503.20019",
    archivePrefix = "arXiv",
    primaryClass = "hep-ph",
    doi = "10.1016/j.physletb.2025.140079",
    journal = "Phys. Lett. B",
    volume = "872",
    pages = "140079",
    year = "2026"
}

@article{Riembau:2024tom,
    author = "Riembau, Marc and Son, Minho",
    title = "{One-point correlators of conserved and nonconserved charges in QCD}",
    eprint = "2407.12082",
    archivePrefix = "arXiv",
    primaryClass = "hep-ph",
    reportNumber = "CERN-TH-2024-113",
    doi = "10.1103/PhysRevD.111.014004",
    journal = "Phys. Rev. D",
    volume = "111",
    number = "1",
    pages = "014004",
    year = "2025"
}

@article{Nielsen:1983rb,
    author = "Nielsen, Holger Bech and Ninomiya, Masao",
    title = "{ADLER-BELL-JACKIW ANOMALY AND WEYL FERMIONS IN CRYSTAL}",
    reportNumber = "BROWN-HET-501",
    doi = "10.1016/0370-2693(83)91529-0",
    journal = "Phys. Lett. B",
    volume = "130",
    pages = "389--396",
    year = "1983"
}

@inproceedings{Kirilin:2013fqa,
    author = "Kirilin, V. P. and Sadofyev, A. V. and Zakharov, V. I.",
    title = "{Anomaly and long-range forces}",
    booktitle = "{100th anniversary of the birth of I.Ya. Pomeranchuk}",
    eprint = "1312.0895",
    archivePrefix = "arXiv",
    primaryClass = "hep-th",
    doi = "10.1142/9789814616850_0014",
    pages = "272--286",
    year = "2014"
}

@article{Coriano:2023hts,
    author = "Corian{\`o}, Claudio and Lionetti, Stefano and Maglio, Matteo Maria",
    title = "{Parity-odd 3-point functions from CFT in momentum space and the chiral anomaly}",
    eprint = "2303.10710",
    archivePrefix = "arXiv",
    primaryClass = "hep-th",
    doi = "10.1140/epjc/s10052-023-11661-1",
    journal = "Eur. Phys. J. C",
    volume = "83",
    number = "6",
    pages = "502",
    year = "2023"
}

@article{Coriano:2024ive,
    author = "Corian{\`o}, Claudio and Cret{\`\i}, Mario and Lionetti, Stefano and Melle, Dario and Tommasi, Riccardo",
    title = "{Axion-Like Interactions and CFT in Topological Matter, Anomaly Sum Rules and the Faraday Effect}",
    eprint = "2403.15641",
    archivePrefix = "arXiv",
    primaryClass = "hep-ph",
    doi = "10.1002/apxr.202400043",
    journal = "Adv. Phys. Res.",
    volume = "4",
    number = "7",
    pages = "2400043",
    year = "2025"
}

@article{Frishman:1980dq,
    author = "Frishman, Y. and Schwimmer, A. and Banks, Tom and Yankielowicz, S.",
    title = "{The Axial Anomaly and the Bound State Spectrum in Confining Theories}",
    reportNumber = "WIS-80/27-Ph",
    doi = "10.1016/0550-3213(81)90268-6",
    journal = "Nucl. Phys. B",
    volume = "177",
    pages = "157--171",
    year = "1981"
}

@article{Landsteiner:2016led,
    author = "Landsteiner, Karl",
    title = "{Notes on Anomaly Induced Transport}",
    eprint = "1610.04413",
    archivePrefix = "arXiv",
    primaryClass = "hep-th",
    reportNumber = "IFT-UAM-CSIC-16-103",
    doi = "10.5506/APhysPolB.47.2617",
    journal = "Acta Phys. Polon. B",
    volume = "47",
    pages = "2617",
    year = "2016"
}

@article{Alipour-fard:2025dvp,
    author = "Alipour-fard, Samuel and Waalewijn, Wouter J.",
    title = "{Energy correlators beyond angles}",
    eprint = "2501.17218",
    archivePrefix = "arXiv",
    primaryClass = "hep-ph",
    reportNumber = "MIT-CTP/5828",
    doi = "10.1007/JHEP07(2025)043",
    journal = "JHEP",
    volume = "07",
    pages = "043",
    year = "2025"
}

@article{Monni:2025zyv,
    author = "Monni, Pier Francesco and Vita, Gherardo and Xu, Zhen and Zhu, Hua Xing",
    title = "{On the Edge of Safety: Charge-Charge Correlation in the Back-to-Back Limit}",
    eprint = "2508.00977",
    archivePrefix = "arXiv",
    primaryClass = "hep-ph",
    reportNumber = "CERN-TH-2025-150",
    month = "8",
    year = "2025"
}

@article{Kimura:1969iwz,
    author = "Kimura, T.",
    title = "{Divergence of axial-vector current in the gravitational field}",
    doi = "10.1143/PTP.42.1191",
    journal = "Prog. Theor. Phys.",
    volume = "42",
    pages = "1191--1205",
    year = "1969"
}

@article{Erdmenger:1999xx,
    author = "Erdmenger, Johanna",
    title = "{Gravitational axial anomaly for four-dimensional conformal field theories}",
    eprint = "hep-th/9905176",
    archivePrefix = "arXiv",
    doi = "10.1016/S0550-3213(99)00561-1",
    journal = "Nucl. Phys. B",
    volume = "562",
    pages = "315--329",
    year = "1999"
}

@article{Coriano:2025ceu,
    author = "Corian{\`o}, Claudio and Lionetti, Stefano and Melle, Dario",
    title = "{Topological sum rules and spectral flows of chiral and gravitational axionlike interactions}",
    eprint = "2502.03182",
    archivePrefix = "arXiv",
    primaryClass = "hep-ph",
    doi = "10.1103/qd7t-4dhl",
    journal = "Phys. Rev. D",
    volume = "112",
    number = "1",
    pages = "015031",
    year = "2025"
}

@article{Coleman:1982yg,
    author = "Coleman, Sidney R. and Grossman, Bernard",
    title = "{'t Hooft's Consistency Condition as a Consequence of Analyticity and Unitarity}",
    reportNumber = "HUTP-82/A009",
    doi = "10.1016/0550-3213(82)90028-1",
    journal = "Nucl. Phys. B",
    volume = "203",
    pages = "205--220",
    year = "1982"
}

@article{Tarasov:2020cwl,
    author = "Tarasov, Andrey and Venugopalan, Raju",
    title = "{Role of the chiral anomaly in polarized deeply inelastic scattering: Finding the triangle graph inside the box diagram in Bjorken and Regge asymptotics}",
    eprint = "2008.08104",
    archivePrefix = "arXiv",
    primaryClass = "hep-ph",
    doi = "10.1103/PhysRevD.102.114022",
    journal = "Phys. Rev. D",
    volume = "102",
    number = "11",
    pages = "114022",
    year = "2020"
}

@book{DonGolHol_SM, 
	place={Cambridge}, 
	title={Dynamics of the Standard Model}, 
	DOI={10.1017/9781009291033}, 
	publisher={Cambridge Univ. Press}, 
	author={Donoghue, John F. and Golowich, Eugene and Holstein, Barry R.}, 
	year={2014}}

@article{Tarasov:2021yll,
    author = "Tarasov, Andrey and Venugopalan, Raju",
    title = "{The role of the chiral anomaly in polarized deeply inelastic scattering II: Topological screening and transitions from emergent axion-like dynamics}",
    eprint = "2109.10370",
    archivePrefix = "arXiv",
    primaryClass = "hep-ph",
    doi = "10.1103/PhysRevD.105.014020",
    journal = "Phys. Rev. D",
    volume = "105",
    number = "1",
    pages = "014020",
    year = "2022"
}

@article{Bhattacharya:2022xxw,
    author = "Bhattacharya, Shohini and Hatta, Yoshitaka and Vogelsang, Werner",
    title = "{Chiral and trace anomalies in deeply virtual Compton scattering}",
    eprint = "2210.13419",
    archivePrefix = "arXiv",
    primaryClass = "hep-ph",
    doi = "10.1103/PhysRevD.107.014026",
    journal = "Phys. Rev. D",
    volume = "107",
    number = "1",
    pages = "014026",
    year = "2023"
}

\end{document}